\documentclass{article}
\usepackage{amssymb}
\usepackage{amsfonts}
\usepackage{multicol}
\usepackage{float}
\usepackage{amsmath}
\usepackage{bm}
\usepackage{cite}
\usepackage[notref,notcite]{showkeys}
\usepackage[dvips]{color}
\usepackage{graphicx}
\usepackage{pstricks}
\allowdisplaybreaks
\psset{unit=0.8cm}
\numberwithin{equation}{section}
\DeclareMathOperator{\diag}{\rm diag}

\DeclareMathOperator{\Rs}{\mathbb{R}}

\DeclareMathOperator{\Zs}{\mathbb{Z}}

\DeclareMathOperator{\No}{\mathcal{N}}
\DeclareMathOperator{\Do}{\mathcal{D}}

\DeclareMathOperator{\Vo}{\mathcal{V}}

\newtheorem{theorem}{Theorem}[section]
\newtheorem{lemma}{Lemma}[section]

\newtheorem{definition}{Definition}[section]

\begin{document}

\title{Properties of the solitonic potentials of the heat operator}
\author{M.~Boiti${}^{*}$, F.~Pempinelli${}^{*}$, and A.~K.~Pogrebkov$
{}^{\dag}$ \\
${}^{*}$Dipartimento di Fisica, Universit\`a del Salento and\\
Sezione INFN, Lecce, Italy\\
${}^{\dag}$Steklov Mathematical Institute, Moscow, Russia}
\date{PACS: 02.30Ik, 02.30Jr, 05.45Yv}
\maketitle

\begin{abstract}
Properties of the pure solitonic $\tau$-function and potential of the heat equation are studied in detail. We describe the asymptotic behavior of the potential and identify the ray structure of this asymptotic behavior on the $x$-plane in dependence on the parameters of the potential.
\end{abstract}

\section{Introduction}

The Kadomtsev--Petviashvili (KP) equation was derived as a model for small-amplitude, long-wavelength, weakly two-dimensional waves in a weakly dispersive medium~\cite{KP1970}. This equation is a (2+1)-dimensional generalization of the celebrated Korteweg--de~Vries (KdV) equation, and from the beginning of the 1970s~\cite{D1974,ZS1974} it is known to be integrable. There are two inequivalent versions of the KP equation: KPI and KPII. Here we consider KPII equation
\begin{equation}
(u_{t}-6uu_{x_{1}}+u_{x_{1}x_{1}x_{1}})_{x_{1}}=-3u_{x_{2}x_{2}}
\label{KPII}
\end{equation}
(the KPI equation has an opposite sign in the r.h.s.), where $u=u(x,t)$, $x=(x_{1},x_{2})$ and subscripts $x_{1}$, $x_{2}$ and $t$ denote partial derivatives. The KPII equation is integrable since it can be expressed as compatibility condition $[\mathcal{L},\mathcal{T}]=0$ of the Lax pair $\mathcal{L}$ and $\mathcal{T}$, where operator
\begin{equation}
\mathcal{L}(x,\partial_{x})=-\partial_{x_{2}}+\partial_{x_{1}}^{2}-u(x)\label{heatop}
\end{equation}
defines the well known equation of heat conduction, or heat equation for short, and
\begin{equation}
\mathcal{T}(x,\partial_{x},\partial_{t})=\partial_{t}+4\partial_{x_{1}}^{3}-6u\partial_{x_{1}}-3u_{x_{1}}-
3\partial_{x_{1}}^{-1}u_{x_{2}}.\label{Lax}
\end{equation}

The spectral theory of the operator (\ref{heatop}) was developed in \cite{BarYacoov,Lipovsky,Wickerhauser,Grinevich0} in the case of a real potential $u(x)$ rapidly decaying at spatial infinity, which, however, is not the most interesting case, since the KPII equation was just proposed in \cite{KP1970} in order to deal with two dimensional weak transverse perturbation of the one soliton solution of the KdV. In fact, if $u_{1}^{}(t,x_{1}^{})$ obeys KdV, then
$u(t,x_{1}^{},x_{2}^{})=u_{1}^{}(t,x_{1}^{}+\mu x_{2}^{}-3\mu_{}^{2}t)$ solves KPII for an arbitrary constant
$\mu\in\Rs$. In particular, KPII admits a one soliton solution of the form
\begin{equation}
u(x,t)=-\dfrac{(\kappa_{1}-\kappa_{2})^{2}}{2}\text{\textrm{sech}}^{2}\Biggl[\dfrac{\kappa_{1}^{}-
\kappa_{2}^{}}{2}x_{1}+\dfrac{\kappa_{1}^{2}-\kappa_{2}^{2}}{2}x_{2}-2(\kappa_{1}^{3}-\kappa_{2}^{3})t\Biggr],
\label{1-sol}
\end{equation}
where $\kappa_{1}$ and $\kappa_{2}$ are real, arbitrary constants.

A spectral theory of the heat operator (\ref{heatop}) that also includes solitons has to be built. In the case of a potential $u(x)$ rapidly decaying at spatial infinity, according to~\cite{BarYacoov,Lipovsky,Wickerhauser,Grinevich0}, the main tools in building the spectral theory of the operator (\ref{heatop}), as in the one dimensional case, are the integral equations defining the Jost solutions. However, if the potential $u(x)$ does not decay at spatial infinity, as it is the case when line soliton solutions are considered (see, e.g., (\ref{1-sol})), these integral equations are ill-defined and one needs a more general approach. In solving the analogous problem for the nonstationary Schr\"{o}dinger operator, associated to the KPI equation, the extended resolvent approach was introduced~\cite{BPP2006b}. Accordingly, a spectral theory of the KPII equation that also includes solitons has to be investigated using the resolvent approach. In this framework it was possible to develop the
inverse scattering transform for a solution describing one soliton on a generic, rapidly decaying background~\cite{BPPP2002}, and to study the existence of the (extended) resolvent for (some) multisoliton solutions~\cite{BPPP2009}.

However, the general case of $N$ solitons is still open. In particular this is motivated by a complicated asymptotic behavior of  the pure solitonic potential on the $x$-plane. Here we give a detailed description of this behavior in terms of the soliton parameters. The general form of the multisoliton potential was derived in \cite{BPPPr2001a} by means of the rational transformations of the scattering data and in \cite{BPPP2009} by means of the twisting transformation. In \cite{equivKPII} we proved that results of both these constructions coincide one with another and also with the form of the multisoliton potential given in terms of the $\tau$-function by Biondini, Kodama et al, see review article \cite{K} and references therein.

The paper is organized as follows. In Sec.\ 2 we write down the multisoliton potential. They are labeled by the two numbers (topological charges), $N_{a}$ and $N_{b}$, that obey condition
\begin{equation}
N_{a},N_{b}\geq 1,  \label{nanb}
\end{equation}
and showing at large spaces $N_{b}$ ``ingoing'' rays and $N_{a}$ ``outgoing'' rays. By using notations of \cite{equivKPII}, we give for these potentials two dual representations in terms of $\tau$-functions, which depend on
\begin{equation}
\No=N_{a}+N_{b},  \label{Nnanb}
\end{equation}
(so that $\No\geq 2$) real parameters $\kappa_n$ and on a $N_a\times N_b$ constant matrix $d$. In Sec.\ 3 we study the asymptotic behavior at large $x$ of the multisoliton potential in details and show that the round angle at the origin can be divided in $\mathcal{N}$ angular sectors such that along the directions of their bordering rays the potential has constant soliton like behavior, while along directions inside the sectors the potential has an exponential decaying behavior, which we derive explicitly. In particular, for a special subclass of the solitonic potentials we present explicitly the ray structure in terms of the parameters  $\kappa_n$.

\section{Multisoliton potentials}

Let we have $\No$ real parameters
\begin{equation}
\kappa_{1}<\kappa_{2}<\ldots <\kappa_{\No},  \label{kappas}
\end{equation}
and introduce the functions
\begin{equation}
K_{n}^{}(x)=\kappa_{n}^{}x_{1}^{}+\kappa_{n}^{2}x_{2}^{},\quad n=1,\ldots,\No.  \label{Kn}
\end{equation}
Let
\begin{equation}
e^{K(x)}=\diag\{e^{K_{n}(x)}\}_{n=1}^{\No},  \label{eK}
\end{equation}
be a diagonal $\No\times {\No}$-matrix, $\Do$ a $\No\times {N_{b}}$ constant real matrix with at least two nonzero maximal minors, and let $\Vo$ be an ``incomplete Vandermonde matrices,'' i.e., the $N_{b}\times\No$-matrix
\begin{equation}
\Vo=\left(\begin{array}{lll}
1 & \ldots & 1 \\
\vdots &  & \vdots \\
\kappa_{1}^{N_{b}-1} & \ldots & \kappa_{\No}^{N_{b}-1}
\end{array}\right) .  \label{W}
\end{equation}
Then, the soliton potential is given by
\begin{equation}
u(x)=-2\partial_{x_{1}}^{2}\log \tau (x),  \label{ux}
\end{equation}
where the $\tau$-function can be expressed as
\begin{equation}
\tau (x)=\det \bigl(\Vo e^{K(x)}\Do\bigr),  \label{tau}
\end{equation}
see the review paper \cite{K}, references therein, and \cite{equivKPII} where the same notations have been used.

There exists (see \cite{BC2}, \cite{BPPP2009}, and \cite{equivKPII}) a dual representation for the potential in terms of the $\tau$-function
\begin{equation}
\tau'(x)=\det \left(\Do^{\,\prime}e^{-K(x)}\gamma \Vo^{\,\prime}\right) ,  \label{tau'}
\end{equation}
where $\Do^{\,\prime}$ is a constant real  $N_{a}\times\No$-matrix that like the matrix $\Do$ has at least two nonzero maximal minors and that is orthogonal to the matrix $\Do$ in the sense that
\begin{equation}
\Do^{\,\prime}\Do=0,  \label{d12}
\end{equation}
being the zero in the r.h.s.\ a $N_{a}\times {N_{b}}$-matrix, and where $\Vo^{\,\prime}$
is the $\No\times {N_{b}}$-matrix
\begin{equation}
\Vo^{\,\prime}=\left(\begin{array}{lll}
1 & \ldots & \kappa_{1}^{N_{a}-1} \\
\vdots &  & \vdots \\
1 & \ldots & \kappa_{\No}^{N_{a}-1}\end{array}\right) ,  \label{sol'}
\end{equation}
and $\gamma$ the constant, diagonal, real $\No\times\No$-matrix
\begin{equation}
\gamma=\diag\{\gamma_n\}_{n=1}^{\No},\qquad
\gamma_n=\prod_{n'=1,n'\neq{n}}^{\No}(\kappa_{n}-\kappa_{n'})^{-1}.\label{gamma}
\end{equation}
This dual representation follows if one notices that
\begin{equation}
\tau (x)=(-1)^{N_{a}N_{b}+N_{a}(N_{a}-1)/2}\left(\prod_{n=1}^{\No}e^{K_{n}(x)}\right)
V(\kappa_{1},\ldots,\kappa_{N_{a}+N_{b}})\tau'(x),  \label{tautau}
\end{equation}
where $V$ denotes the Vandermonde determinant
\begin{equation}
V(\kappa_{1},\ldots ,\kappa_{\No})=\prod_{1\leq m<n\leq \No}(\kappa_{n}-\kappa_{m}).  \label{V}
\end{equation}

Matrices $\Do$ and $\Do^{\,\prime}$ obey rather interesting properties \cite{equivKPII}. Taking into account that the matrix $\Do$ has  nonzero maximal minors we can always write it in the form
\begin{equation}
\Do=\pi\left(\begin{array}{c}
d_1\\d_{2}\end{array}\right),\label{block1}
\end{equation}
where $\pi$ is a $\No\times\No$-permutation matrix such that the $N_b\times{N_b}$-matrix $d_2$ is nonsingular. Then, we can rewrite this equality in the form
\begin{equation}
\Do=\pi\left(\begin{array}{c}
d \\E_{N_{b}}\end{array}\right)d_2\label{block}
\end{equation}
where $d=d_{1}^{}d^{-1}_{2}$ is a constant real $N_{a}\times {N_{b}}$-matrix and $E_{N_{b}}$ the $N_{b}\times N_{b}$ unity matrix.

Now, we notice (see \cite{K}) that the expression for the potential in (\ref{ux}), as well as condition (\ref{d12}), are left unchanged if in (\ref{tau}) and (\ref{tau'}), respectively, we multiply the $\No\times {N_{b}}$-matrix $\Do$ from the right by a nonsingular constant $N_b\times{N_b}$-matrix and the $N_{a}\times\No$-matrix $\Do^{\,\prime}$ from the left by a nonsingular constant  $N_a\times{N_a}$-matrix. We conclude that the matrix $d_2$ is unessential and that the matrix $\Do$ can be chosen to have a simple block structure. It is clear, however, that choice of matrixes $\pi$ and $d$ is not unique as the matrix $\Do$ can have different nonzero maximal minors.

Let us write now $\Do'\pi=(d'_{1},d'_{2})$, where $d_1'$ is $N_a\times{N_a}$-matrix and $d'_2$ is  $N_a\times{N_b}$-matrix. Then by (\ref{d12}) we get $d_{2}'=-d'_{1}d$, where  $d$ is the same $N_{a}\times {N_{b}}$ matrix as in~(\ref{block}). Thus, for $\Do'$ we get the block structure
\begin{equation}
\Do^{\,\prime}=d'_{1}( E_{N_{a}},-d)\pi^{\dag},\label{block'}
\end{equation}
and thanks to equivalence of $\tau$ and $\tau'$ we get that $\det{d'_1}\neq0$ and, then, also $\Do^{\,\prime}$ can be chosen to have a simple block structure.

Moreover, by means of the block structures given in (\ref{block}) and  (\ref{block'}) we get
\begin{equation}
{\Do}^{\dag}\Do=d^{\dag}_2(E_{N_{b}}+d^{\dag}d)d_2,\qquad \Do^{\,\prime}{\Do^{\,\prime}}^{\dag}=d'_{1}(E_{N_{a}}+dd^{\dag}){d'_1}^{\dag},  \label{d14}
\end{equation}
where $\dag $ denotes the Hermitian conjugation of a matrix (in fact, transposition here). So both matrices ${\Do}^{\dag}\Do$ and $\Do^{\,\prime}{\Do^{\,\prime}}^{\dag}$ are positive and then $\Do$ and $\Do^{\,\prime}$ admit, respectively, left and right inverse (see~\cite{G1990}). Precisely, we have
\begin{equation}
\bigl(\Do\bigr)^{(-1)}\Do=E_{N_{b}},\qquad \Do^{\,\prime}\bigl(\Do^{\,\prime}\bigr)^{(-1)}=E_{N_{a}},  \label{d17}
\end{equation}
with
\begin{equation}
\bigl(\Do\bigr)^{(-1)}=({\Do}^{\dag}\Do)^{-1}{\Do}^{\dag}, \qquad
\bigl(\Do^{\,\prime}\bigr)^{(-1)}=
{\Do^{\,\prime}}^{\dag}(\Do^{\,\prime}{\Do^{\,\prime}}^{\dag})^{-1}.  \label{d15}
\end{equation}
Products of these matrices in the opposite order give the real Hermitian $\No\times\No$-matrices
\begin{align}
& P=\Do\bigl(\Do\bigr)^{(-1)}=\Do({\Do}^{\dag}\Do)^{-1}\bigl(\Do\bigr)^{\dag},  \label{d19} \\
& P^{\,\prime}=\bigl(\Do^{\,\prime}\bigr)^{(-1)}\Do^{\,\prime}=
\bigl(\Do^{\,\prime}\bigr)^{\dag}(\Do^{\,\prime}{\Do^{\,\prime}}^{\dag})^{-1}\Do^{\,\prime},\label{d18}
\end{align}
which are orthogonal projectors, i.e.,
\begin{equation}
P^{2}=P,\qquad (P^{\,\prime})^{2}=P^{\,\prime},\qquad PP^{\,\prime}=0=P^{\,\prime}P,  \label{d22}
\end{equation}
and complementary in the sense that
\begin{equation}
P+P^{\,\prime}=E_{\No}.  \label{d23}
\end{equation}
Orthogonality of the projectors follows from~(\ref{d12}) and the last equality from obvious relations of the kind
$(E_{N_{b}}+d^{\dag}d)^{-1}d^{\dag}=d^{\dag}(E_{N_{a}}+dd^{\dag})^{-1}$.

Finally, let us mention that if all maximal minors of the matrix $\Do$ are nonzero, the permutation $\pi$ in (\ref{block}) and (\ref{block'}) can be arbitrarily chosen. Then, we can chose it equal to the identity, getting the simplified representations
\begin{equation}
\Do=\left(\begin{array}{c}
d \\E_{N_{b}}\end{array}\right)d_2,\qquad
\Do^{\,\prime}=d'_{1}( E_{N_{a}},-d),\label{blockreg}
\end{equation}
so that in this special case without loss of generality we choose
\begin{equation}
\Do=\left(\begin{array}{c}
d \\E_{N_{b}}\end{array}\right),\qquad
\Do^{\,\prime}=( E_{N_{a}},-d).\label{block2}
\end{equation}

In order to study the properties of the potential, it is convenient to use an explicit representation for the
determinant (\ref{tau}). By using the Binet--Cauchy formula for the determinant of a product of matrices we get
\begin{equation}
\tau (x)=\sum_{1\leq n_{1}<n_{2}<\cdots <n_{N_{b}}\leq {\No}}f_{n_{1},\ldots,n_{N_{b}}}
\prod_{l=1}^{N_{b}}e^{K_{n_{l}}(x)},\label{tauf}
\end{equation}
with coefficients $f_{n_{1},n_{2},\ldots ,n_{N_{b}}}$ given by
\begin{equation}
f_{n_{1},n_{2},\ldots ,n_{N_{b}}}=V(\kappa_{n_{1}}^{},\ldots ,\kappa_{n_{N_{b}}}^{})\Do(n_{1},\ldots ,n_{N_{b}}). \label{f}
\end{equation}
Here we used notation (\ref{V}) for the Vandermonde determinant and notation
\begin{equation}
\Do(n_{1},\ldots ,n_{N_{b}})=\det \left(\begin{array}{lll}
\Do_{n_{1},1} & \dots & \Do_{n_{1},N_{b}} \\
\vdots &  & \vdots \\
\Do_{ n_{N_{b}},1} & \dots & \Do_{n_{N_{b}},N_{b}}\end{array}\right) ,\label{Do}
\end{equation}
for the maximal minors of the matrix $\Do$. Notice that these coefficients are invariant
under permutations of the indices.

From (\ref{tauf}) it follows directly that condition
\begin{equation}
f_{n_{1},\ldots ,n_{N_{b}}}\geq 0,\quad \text{for all }1\leq n_{1}<n_{2}<\cdots
<n_{N_{b}}\leq {\No},  \label{reg}
\end{equation}
is sufficient (see \cite{K}) for the regularity of the potential $u(x)$, i.e., for the absence of zeros of $\tau(x)$ on the $x$-plane. Thanks to (\ref{kappas}), (\ref{V}) and (\ref{f}) this condition is equivalent to the condition that all maximal minors of the matrix $\Do$ are non negative. In \cite{equivKPII} it was mentioned that condition (\ref{reg}) is also necessary for the regularity of a  potential under evolution with respect to an arbitrary number of higher times of the KP hierarchy. In \cite{K} it is suggested to decompose soliton solutions of KPII into subclasses, associated to the Schubert cell on Grassmanian. It is proved there that condition (\ref{reg}) is necessary for the regularity of the all solutions associated to a cell. However, problem of finding of the conditions necessary for the regularity of the many soliton solution under evolution with respect to KPII only is still opened.

If the matrix $\Do$ can be chosen as in (\ref{block2}), then introducing new $N_a\times{N_b}$-matrix $\widetilde{d}$ as
\begin{equation}
\widetilde{d}_{n,l}=d_{N_a+1-n,l}(-1)^{l+1}, \quad n=1,\ldots,N_a,\quad l=1,\ldots,N_b,\label{dd2}
\end{equation}
we easily get by (\ref{Do}) that
\begin{align}
&\Do(n_{1},\ldots ,n_{N_{b}})=\det\Vert{\widetilde{d}_{n_i,l}}\Vert,\nonumber\\
&\text{where} i=1,\ldots,k,\quad l=1,\ldots,N_b,\quad
l\neq n_j-N_a,\quad j=k+1,\ldots,N_b,\label{dd1}
\end{align}
and where number $k$, $1\leq{k}\leq{N_b}$ is defined by a condition $n_k\leq{N_a}<n_{k+1}$. This proves that if we require condition (\ref{reg}) on coefficients $f_{n_{1},\ldots ,n_{N_{b}}}$, then the matrix $\widetilde{d}$ is a totally nonegative (positive, if all inequalities in (\ref{reg}) are strict) one~(see (\cite{ando, G2002}).

In what follows it is convenient to continue indexes of $\kappa_{n}$, $f_{n_1,\ldots,n_{N_b}}$, etc, periodically by $\No$ on the whole $\Zs$ by means of condition:
\begin{equation}
n\to n\,(\text{mod}\No).  \label{modN}
\end{equation}
Then Eq. (\ref{tauf}) can be written as
\begin{equation}
\tau (x)=\sum_{n\leq n_{1}<n_{2}<\cdots <n_{N_{b}}\leq {\No+n-1}}f_{n_{1},\ldots,n_{N_{b}}}
\prod_{l=1}^{N_{b}}e^{K_{n_{l}}(x)},\label{taufn}
\end{equation}
where the r.h.s.\ is independent of $n$, and where coefficients  $f_{n_1,\ldots,n_{N_b}}$ are defined by (\ref{f}) taking (\ref{modN}) into account. Thanks to (\ref{kappas}) and (\ref{modN}) we can prove the following lemma.
\begin{lemma}\label{lemma1} Let
\begin{equation}
g_{l,m,n}=(\kappa_{l}-\kappa_{m})(\kappa_{n}-\kappa_{n+N_b})(\kappa_{l}+\kappa_{m}-\kappa_{n}-\kappa_{n+N_b}).
\label{l11}
\end{equation}
Then
\begin{equation}
g_{l,m,n}\geq0\quad\text{for any}\quad n\in\Zs,\quad l=n,\ldots,n+N_b,\quad m=n+N_b,\ldots,\No+n,\label{l12}
\end{equation}
and equality takes place only when $l$ and $m$ independently take values $n$ or $n+N_b$ by mod$\,\No$.
\end{lemma}

{\sl Proof.\/} Thanks to (\ref{modN}) it is enough to prove the lemma for $n=1,\ldots,\No$. For these values of $n$ the $\kappa_{n}$'s are ordered as in (\ref{kappas}). For other values the ordering is obtained by coming back by means of  (\ref{modN}) to values belonging to the interval $n=1,\ldots,\No$. Thus, when $n=1,\ldots,N_a$ then $n$, $n+N_b$ and $l$ are not more than $\No$ (see (\ref{Nnanb})), but instead of the interval for $m$ given in  (\ref{l12}) we get two intervals: $m=n+N_b,\ldots,\No$ and $m=1,\ldots,n$. Then for given values of $n$, $l$ and $m$ in the first interval lemma follows because the first factor in (\ref{l11}) is nonpositive, the second one negative and the third one nonnegative. Zero of the first factor is possible only for $l=m$ and then $l=m=n+N_b$ and of the third factor only when both $l=n$ and $m=n+N_b$ (all equalities by mod$\,\No$). In the second case the first factor is nonnegative (zero only at $l=m=n$~(mod$\,\No$)), the second one is negative and the third one is nonpositive (zero only at $l=n+N_b$, $m=n$). In the interval $n=N_a+1,\ldots,\No$ we write by (\ref{modN}) $\kappa_{n+N_b}=\kappa_{n-N_a}$, so that again $n$ and $n-N_a$ belong to the interval $1,\ldots,\No$. Now the interval of $l$ given in (\ref{l12}) decouples in two intervals $n,\ldots,\No$ and $1,\ldots,n-N_a$, while $m=n-N_a,\ldots,n$, so it is positive and less than $\No$. Then for these values of $n$, $m$ and $l$ in the first interval thanks to (\ref{kappas}) the first factor in (\ref{l11}) is nonnegative (zero only when $l=m=n$~(mod$\,\No$)), the second factor is positive, and the third one is nonnegative (zero only when $l=n$ and $m=n+N_b$). In the case of the second interval the first factor is nonpositive (zero only when $l=m=n-N_a=n+N_b$~(mod$\,\No$)), the second factor is positive and the third factor is nonpositive (zero only when $l=n-N_a=n+N_b$~(mod$\,\No$) and $m=n$). $\blacksquare$

\section{Asymptotic behavior of the $\tau$-function and potential $u(x)$}

\begin{figure}[tbp]
\begin{center}
\begin{pspicture}[showgrid=false,linewidth=1.0pt](-4.5,-4.5)(4.5,4.5)
%\psgrid[gridwidth=0.2pt,subgridwidth=0.2pt]
\psline[linestyle=dashed]{->}(0,-4.3)(0,4.3)
\psline[linestyle=dashed]{->}(-4.3,0)(4.3,0)
\psline(3,4)(0,0)(1,4)
\psline(4,-1)(0,0)(4,0.5)
\psline(-4,-1.25)(0,0)(-4,0.3)
\psline[linestyle=dotted]{<->}(0.75,3.1)(1.4,3)(2,2.7)
\psline[linestyle=dotted]{<->}(-2.6,-0.8)(-2.85,-0.35)(-2.8,0.2)
\psline[linestyle=dotted]{<->}(2.8,-0.7)(2.95,-0.2)(2.75,0.32)
\rput(-3.2,-0.4){\small$\sigma^{}_{1}$}
\rput(1.5,3.2){\small$\sigma^{}_{n}$}
\rput(3.35,-0.35){\small$\sigma^{}_{N_a}$}
\uput[0](4.4,-0.4){$x^{}_1$}
\uput[90](-0.3,4.1){$x^{}_2$}
\uput[0](-3.6,-1.4){\small$r^{}_{1}$}
\uput[0](3.5,-1.2){\small$r^{}_{N_a-1}$}
\uput[0](3.5,0.75){\small$r^{}_{N_a}$}
\uput[0](3.5,3.75){\small$r^{}_{n-1}$}
\uput[0](0.7,3.75){\small$r^{}_{n}$}
\uput[0](-3.6,.8){\small$r^{}_{\No}$}
\end{pspicture}
\end{center}
\caption{Sector $\protect\sigma_{n}$ corresponds to $N_{a}+2\leq n\leq\No-1$.}
\label{fig}
\end{figure}
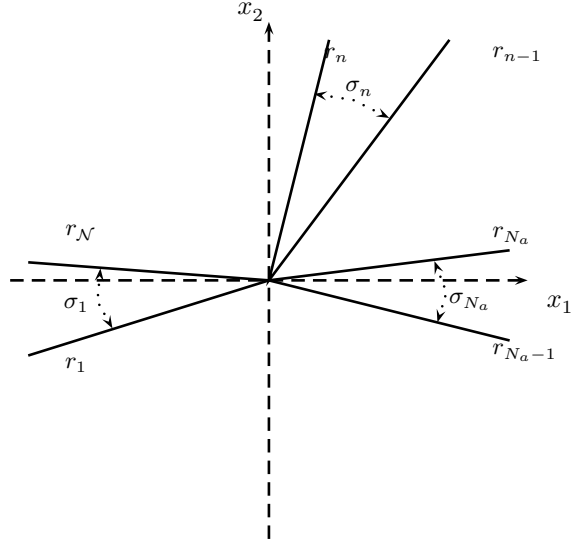

Here we study the asymptotic behavior of the function $\tau(x)$ that by (\ref{tauf}) is determined by interrelations between functions $K_{n}(x)$ for different $n$. Since by (\ref{Kn})
\begin{equation}
K_{l}(x)-K_{m}(x)=(\kappa_{l}-\kappa_{m})(x_{1}+(\kappa_{l}+\kappa_{m})x_{2}),\qquad
\text{for any }l,m\in \mathbb{Z},\label{linen}
\end{equation}
that is, these differences are linear with respect to the space variables, the asymptotic behavior must have sectorial structure on the $x$-plane. In order to describe this structure we introduce rays $r_{n}$ given by
\begin{equation}
r_{n}=\{x:x_{1}+(\kappa_{n+N_{b}}+\kappa_{n})x_{2}=0,\,(\kappa_{n+N_{b}}-\kappa_{n})x_{2}<0),\qquad
n=1,\ldots ,\No,  \label{rayn}
\end{equation}
that intersect the lines $x_{2}=\pm 1$, respectively, at the points
\begin{align}
&(\kappa_{n+N_b}+\kappa_n,-1) \qquad \text{for\ } n=1,\dots,N_a\notag\\
&(-\kappa_{n+N_b}-\kappa_n,1) \qquad \text{for\ } n=N_a+1,\dots,\No.\label{y0n}
\end{align}
Thanks to (\ref{kappas}) and (\ref{modN}),
\begin{align}
&\kappa_{n+N_b}-\kappa_{n}> 0, &n = 1,\ldots, N_{a} , \label{ineq1}\\
&\kappa_{n+N_b}-\kappa_{n}<0, & n = N_{a} +1,\ldots,\No,\label{ineq2}
\end{align}
therefore, increasing $n$ from $1$ to $N_{a}$ the ray rotates anticlockwise in the lower half $x$-plane, crosses the positive part of the $x_{1}$-axis in coming to $n=N_{a}+1$ and, then, rotates anticlockwise in the upper half $x$-plane up to $n=\No$ and, finally, crossing the negative part of the $x_{1}$-axis, for $n=\No+1$ comes back to the ray $r_{1}$ thanks to (\ref{modN}), see Fig.~\ref{fig}.

Let us assume that some rays, say, $r_{m}$ and $r_{n}$, where for definiteness $m<n$, are parallel. By (\ref{rayn}) this means that $\kappa_{m+N_{b}}+\kappa_{m}=\kappa_{n+N_{b}}+\kappa_{n}$, i.e., that $\kappa_{n}-\kappa_{m}=\kappa_{m+N_{b}}-\kappa_{n+N_{b}}$, where the l.h.s.\ is positive thanks to (\ref{kappas}). Then, because of (\ref{modN}) it is easy to see that the r.h.s.\ can be positive only if $1\leq{m}\leq{N_a}<{n}\leq\No$, so that by (\ref{rayn}) ray $r_m$ is in the bottom halfplane and $r_n$ in the upper one. Thus rays can be parallel only if they belong to different halfplanes, but in the case $N_a\neq N_b$ this is possible for a special choice of the parameters $\kappa_{n}$ only. On the contrary, in the case $N_a=N_b$ all pairs $r_{n}$ and $r_{n+N_a}$, $n=1,\ldots,N_a$, and only these pairs give parallel rays. In the special case $N_{a}=N_{b}=1$ we get two rays producing the straight line  $x_{1}+(\kappa_{1}+\kappa_{2})x_{2}=0$ that divides the $x$-plane in two halfplanes.

Now we introduce sectors $\sigma_n$, which are subsets of the $x$-plane characterized as
\begin{equation}
\sigma_{n}=\{x:K_{n-1}(x)<K_{n+N_{b}-1}(x)\text{ and }K_{n}(x)>K_{n+N_{b}}(x)\},\quad
n=1,\dots ,\mathcal{N}.\label{sigman}
\end{equation}
Thanks to (\ref{linen}) and the discussion above it follows that the sectors $\sigma_{n}$ are sharp (for $\No>2$) angular sectors with vertexes at the origin of the coordinates bounded from the right (looking from the origin) by the ray $r_{n-1}$ and from the left (looking from the origin) by the ray $r_{n}$. With the increasing of $n$ sectors $\sigma_{n}$ are ordered anticlockwise, starting ``from the left'' with the sector $\sigma_{1}$, that includes the negative part of the $x_{1}$-axis, and, then, with the sectors $\sigma_{n}$ ($n=2,\ldots ,N_{a}$) in the bottom half-plane, the sector $\sigma_{N_{a}+1}$ ``to the right'', that includes the positive part of the $x_{1}$-axis, and the sectors $\sigma_{n}$ ($n=N_{a}+2,\ldots ,\No$) on the upper half-plane, finishing with the sector $\sigma_{\No}$ tangent to the sector $\sigma_{1}$, covering in this way the whole $x$-plane with the exception of the bordering rays $r_n$. Therefore, the sectors $\sigma_{n}$ define a $\mathcal{N}$-fold discretization of the round angle at the origin, with $n$ playing the role of a discrete angular variable, see Fig.~\ref{fig}. It is clear that in the study of asymptotic behavior of $\tau$-function we consider $x$-plane as vector space, as finite part of $x$ when $x\to\infty$ is irrelevant. For determining the directions of rays $r_{n}$ and sectors $\sigma_{n}$, we introduce the vectors
\begin{equation}
y_{n}=(\kappa_{n+N_{b}}^{2}-\kappa_{n}^{2},\kappa_{n}-\kappa_{n+N_{b}}),\qquad n=1,\dots ,\No,  \label{pointn'}
\end{equation}
that enables us to give the following definition.

\begin{definition}\label{def1'}
We say that $x\to\infty$ along the ray $r_n$ if $x\to\infty$ and there exists such function $\alpha(x)\to+\infty$ that $x-\alpha(x)y_{n}$ is bounded. This will be denoted as $x\stackrel{r_{n}}{\longrightarrow}\infty$.

We say that $x\to\infty$ in the sector $\sigma_{n}$, $n=1,\ldots,\No$, if $x\to\infty$ and there exist such functions $\alpha(x)\to+\infty$ and $\beta(x)\to+\infty$ that $x-\alpha(x)y_{n-1}-\beta(x)y_{n}$ is bounded. This will be denoted as $x\stackrel{\sigma_{n}}{\longrightarrow}\infty$.
\end{definition}

Notice that for  $x\stackrel{r_{n}}{\longrightarrow}\infty$
\begin{equation}\label{def12}
K_{n}(x)-K_{n+N_b}(x)\quad\text{is bounded and}\qquad (\kappa_{n+N_b}-\kappa_{n})x_2\to-\infty,
\end{equation}
and for $x\stackrel{\sigma_{n}}{\longrightarrow}\infty$
\begin{equation}\label{def11}
K_{n+N_b-1}(x)-K_{n-1}(x)\to+\infty,\qquad K_{n}(x)-K_{n+N_b}(x)\to+\infty,
\end{equation}
as follows directly from the definition. In fact we can prove a more general statement.
\begin{lemma}
\label{lemma3} Let $N_{a},N_{b}\geq 1$, and $n\in\Zs$ arbitrary. Then
\begin{enumerate}
\item if $x\stackrel{r_{n}}{\longrightarrow}\infty$ then $K_{l}(x)-K_{m}(x)\to+\infty$ or bounded for any $l=n,\ldots,n+N_{b}$ and $m=n+N_{b},\ldots,\No+n$, where boundedness takes place if and only if $(l,m)=(n,n)$, $(n,n+N_b)$, $(n+N_b,n)$, $(n+N_b,n+N_b)$;
\item if $x\stackrel{\sigma_{n}}{\longrightarrow}\infty$ then $K_{l}(x)-K_{m}(x)\to+\infty$ for any
$l=n,\ldots,n+N_{b}-1$ and $m=n+N_{b},\ldots,\No+n-1$;
\end{enumerate}
where summation in indexes is always understood by mod$\,\No$.
\end{lemma}
\textsl{Proof.\/} Thanks to (\ref{linen}) and (\ref{pointn'}) we have that $K_{l}(y_{n})-K_{m}(y_n)=g_{l,m,n}$. Then the first statement of the lemma follows directly from Definition \ref{def1'} and Lemma \ref{lemma1}. In the same way we get that $K_{l}(x)-K_m(x)=\alpha g_{l,m,n-1}+\beta g_{l,m,n}$ under substitution $x=\alpha y_{n-1}+\beta y_n$. Then the second statement of the lemma follows from the second statement of the Lemma \ref{lemma1} while intervals for $l$ and $m$ appear as intersections of the corresponding intervals in (\ref{l12}) for $n$ and $n\to{n-1}$. $\blacksquare $

Now we can prove the following Theorem.
\begin{theorem}
\label{th1} The asymptotic behavior of $\tau(x)$ for $x\rightarrow\infty$ is given by
\begin{align}
&x\stackrel{r_{n}}{\longrightarrow}\infty:&&
\tau (x)=\bigl(z_{n}+z_{n+1}e_{}^{K_{N_{b}+n}(x)-K_{n}(x)}+o(1)\bigr)\exp\left(\sum_{j=n}^{n+N_{b}-1}K_{j}(x)
\right), \label{3:232}\\
&x\stackrel{\sigma_{n}}{\longrightarrow}\infty:&&\tau (x)=\bigl(z_{n}+o(1)\bigr)\exp\left(\sum_{l=n}^{n+N_{b}-1}K_{l}(x)\right),  \label{tauasympt}
\end{align}
for any $n\in\mathbb{Z}$, where coefficients $z_n$ are defined as
\begin{equation}
z_{n}=f_{n,n+1,\ldots ,n+N_{b}-1}\equiv V(\kappa _{n}^{},\ldots ,\kappa_{n+N_{b}-1}^{})\Do(n,\ldots ,n+N_{b}-1),  \label{zn}
\end{equation}
(cf.\ (\ref{f})). Terms $o(1)$ are decaying exponentially.
\end{theorem}

\textsl{Proof.\/} Using representation~(\ref{taufn}) we can write
\begin{align}
&\exp\left(-\sum_{l=n}^{n+N_{b}-1}K_{l}(x)\right) \tau (x)=\nonumber\\
&\qquad=\sum_{n\leq n_{1}<n_{2}<\cdots <n_{N_{b}}\leq {\No+n-1}}f_{n_{1},\ldots ,n_{N_{b}}}\exp\left( \sum_{j=1}^{N_{b}}K_{n_{j}}(x)-\sum_{l=n}^{n+N_{b}-1}K_{l}(x)\right) .
\end{align}
Let us consider first the asymptotics when $x\stackrel{\sigma_{n}}{\longrightarrow}\infty$. Then each term $K_{n_{j}}(x)$ either cancels with some term in the last sum, when the index $n_{j}$ equals by mod\,$\No$ one index $l$ in the interval $\{n,\ldots ,n+N_{b}-1\}$, i.e., an interval of the index $l$ in Lemma~\ref{lemma3}, or belong to the interval $\{n+N_{b},\ldots ,\No+n-1\}$. Thus by Lemma~\ref{lemma3} all these exponents at infinity are negative with the only exception of the case where $\{n_{1},\ldots ,n_{N_{b}}\}=\{n,\ldots ,n+N_{b}-1\}(\text{mod}\No)$. This proves  (\ref{tauasympt}). Proof of (\ref{3:232}) goes in the same way with the only difference that we use the first statement of the Lemma~\ref{lemma3}. Statement on the asymptotic behavior of the terms of $o(1)$ kind is obvious by construction. $\blacksquare $

As we see from (\ref{3:232}) and (\ref{tauasympt}) the leading asymptotic behavior is fixed by this theorem only if all
\begin{equation}
z_{n}>0,\qquad n=1,\ldots ,\No. \label{zn0}
\end{equation}
This condition is a sufficient for the regularity of the potential at large $x$, while specific examples show that it is not sufficient for the regularity of the potential in all the $x$-plane. We also see that the asymptotics along the ray direction $r_{n}$ is given by the sum of the leading terms obtained for $x\stackrel{\sigma_{n}}{\longrightarrow}\infty$ and  $x\stackrel{\sigma_{n+1}}{\longrightarrow}\infty$. Since the exponential factor cancels out when expansion (\ref{3:232}) of $\tau (x)$ is inserted in~(\ref{ux}), the factor in parenthesis gives the ray behavior of the potential $u(x)$ at infinity. Thus, if condition (\ref{zn0}) is imposed potential $u(x)$ has exactly $\No$ rays $r_n$, obeying nontrivial asymptotic behavior of onesoliton potential (cf.\ (\ref{1-sol})) with parameters $\kappa_n$ and $\kappa_{n+N_b}$. Explicitly, taking into account (\ref{modN}) we get the following ray structure and asymptotic behavior of the potential along rays:
\begin{align}
&x_{2}\rightarrow-\infty,\qquad x_{1}+(\kappa_{n}+\kappa_{n+N_{b}})x_{2}\text{ being bounded:}\nonumber\\
&u(x)\cong-2\partial_{x_{1}}^{2}\log\bigl(z_{n}+z_{n+1}e_{}^{K_{n+N_{b}}(x)-K_{n}(x)}\bigr),
\quad  n=1,\ldots ,N_{a}.\label{down}\\
\intertext{and}
&x_{2}\rightarrow+\infty,\qquad x_{1}+(\kappa_{n}+\kappa_{n+N_{a}})x_{2}\text{ being bounded:}\nonumber\\
&u(x)\cong-2\partial_{x_{1}}^{2}\log\bigl(z_{n+N_{a}}+z_{n+N_{a}+1}e_{}^{K_{n}(x)-K_{n+N_{a}}(x)}\bigr),
\quad n=1,\ldots ,N_{b},\label{up}
\end{align}
i.e., $N_a$ asymptotic rays in the bottom half plane and $N_b$ asymptotic rays in the upper half-plane. Asymptotic behavior inside sectors $\sigma_{n}$ is given by (\ref{tauasympt}) where the only $x$-dependent term is the exponential factor.  Taking into account its linear dependence on $x$ and (\ref{zn0}) we get that $u(x)$ decays exponentially in all directions inside all sectors.

\section{Concluding remarks}

Asymptotic behavior of the $\tau$-function and potential derived here is based on condition~(\ref{zn0}) allowing vanishing or negative values of coefficients in (\ref{tauf}) different from $z_n$ (see (\ref{zn})). Thanks to~(\ref{zn0}) we were able to find explicitly that the ray directions are given by the sums $\kappa_{n}+\kappa_{n+N_b}$, with $n=1,\ldots, \No$, and to derive the exact asymptotic behavior of the $\tau$-function on the $x$-plane. Let us mention that analogous result was derived in \cite{BK} for a very special subclass of solitonic potentials, given by matrix $\Do$ in (\ref{tau}) equal to a product of a diagonal matrix by the matrix transpose to $\Vo$. As we mentioned above, condition (\ref{zn0}) does not guarantee (in contrast to~(\ref{reg})) absence of singularities of the potential $u(x)$. But (\ref{3:232}) and (\ref{tauasympt}) prove that thanks to (\ref{zn0}) singularities of the potential (\ref{ux}) cannot appear in asymptotics, as inside sectors $u(x)\to0$ and along rays $r_n$ it is given by (\ref{down}) and (\ref{up}), that is also finite thanks to  (\ref{zn0}).  Let us emphasize that parameters $N_a$ and $N_b$ giving numbers of rays of the $\tau$-function and potential in the bottom and upper half-planes, correspondingly,  appeared in \cite{BPPP2009} as numbers of poles of the Jost and dual Jost solutions. And, finally, we mention that Theorem \ref{th1} is also valid when all inequalities in the regularity requirement (\ref{reg}) are strict, i.e., all maximal minors of the matrix $\Do$ are positive, as this condition is stronger than (\ref{zn0}) that was enough to prove the theorem.

\section{Acknowledgments}

This work is supported in part by the grant RFBR \# 08-01-00501, grant RFBR--CE \# 09-01-92433, Scientific Schools 8265.2010.1, by the Program of RAS ``Mathematical Methods of the Nonlinear Dynamics,'' by INFN, by the grant PRIN 2008 ``Geometrical methods in the theory of nonlinear integrable systems,'' and by Consortium E.I.N.S.T.E.IN. AKP thanks Department of Physics of the University of Salento (Lecce) for kind hospitality.

\end{document}